\documentclass[aps,prx-quantum,twocolumn,groupedaddress,amsmath,nofootinbib]{revtex4-1}
\usepackage{graphicx}% Include figure files
\usepackage{subfigure}
\usepackage{mathtools}
\usepackage{mathrsfs} 
\usepackage{dcolumn}% Align table columns on decimal point
\usepackage{bm}% bold math
\usepackage{amsmath}
\usepackage{color}
\usepackage{verbatim}
\usepackage{natbib} 
\usepackage{verbatim}
\usepackage{amsfonts}
\usepackage{lipsum}
\usepackage{epstopdf}
\usepackage[colorlinks,
            linkcolor=blue,
            anchorcolor=blue,
            citecolor=blue,
            urlcolor=blue]{hyperref}
\renewcommand{\vec}[1]{\boldsymbol{#1}}
\usepackage{bbm}

\begin{document}

\title{A Geometric Quantum Speed Limit: Theoretical Insights and Photonic Implementation}

\author{Qianyi Wang}
\affiliation{National Laboratory of Solid State Microstructures, Key Laboratory of Intelligent Optical Sensing and Manipulation, 
College of Engineering and Applied Sciences and school of physics, and Collaborative Innovation Center of Advanced Microstructures, Nanjing University, Nanjing 210093, China}

\author{Ben Wang}
\email{ben.wang@nju.edu.cn}
\affiliation{National Laboratory of Solid State Microstructures, Key Laboratory of Intelligent Optical Sensing and Manipulation, 
College of Engineering and Applied Sciences and school of physics, and Collaborative Innovation Center of Advanced Microstructures, Nanjing University, Nanjing 210093, China}

\author{Jun Wang}
%\email{wangj@nju.edu.cn}
\affiliation{National Laboratory of Solid State Microstructures, Key Laboratory of Intelligent Optical Sensing and Manipulation, 
College of Engineering and Applied Sciences and school of physics, and Collaborative Innovation Center of Advanced Microstructures, Nanjing University, Nanjing 210093, China}

\author{Lijian Zhang}
\email{lijian.zhang@nju.edu.cn}
\affiliation{National Laboratory of Solid State Microstructures, Key Laboratory of Intelligent Optical Sensing and Manipulation, 
College of Engineering and Applied Sciences and school of physics, and Collaborative Innovation Center of Advanced Microstructures, Nanjing University, Nanjing 210093, China}
\date{\today}

\begin{abstract}
Quantum mechanics imposes a lower bound on the time required for a quantum system to reach certain given targets.
In this paper, from a geometric perspective, we introduce a new quantum speed limit (QSL) based on the Bloch angle 
and derive the condition for it to saturate. Experimentally, we demonstrate the feasibility of measuring this QSL using a photonic system through direct Bloch angle measurements via a swap test, bypassing the need for comprehensive quantum state tomography. Compared to the existing Bloch-angle-based QSL mentioned in prior work, our QSL requires fewer computational and experimental resources and provides tighter constraints for specific dynamics.
Our work underscores the Bloch angle's effectiveness in providing tighter and experimentally accessible QSLs and advances the understanding of quantum dynamics.
\end{abstract}

\maketitle

\section{Introduction}
Quantum mechanics imposes a limit on the minimal time required for a quantum system to evolve from its initial state to another related state. Quantum speed limits (QSLs), defined as the lower bound of the evolution time, 
have numerous applications in quantum computation \cite{Lloydnature2000}, quantum communication \cite{InformationTransfer}, quantum thermodynamics \cite{Deffner-Lutz-CLausius-inequality,Energy-efficient-quantum-machines,incohent-heat-engine}, quantum battery \cite{quantum-battery-Binder_2015,quantum-battery-campaioli,quantum-battery-juli}, quantum metrology \cite{Giovannetti-Lloyd-Nature-Photonics} and quantum optimal control theory \cite{Optimal-Control-at-the-QSL,ita_quatum_optimal_control,driving_at_the_qsl,Time-Optimal-Quantum-Evolution,Solution-to-the-Quantum-Zermelo-Navigation-Problem,Quantum-Brachistochrone-Curves-as-Geodesics}.
Historically, two well-known QSLs exist: the Mandelstam-Tamm (MT) bound \cite{MT1945} and the Margolus-Levitin (ML) bound \cite{ML1998}. The MT bound quantifies the minimal time required for a quantum system to evolve from an initial pure state to an orthogonal pure state under unitary dynamics governed by a time-independent Hamiltonian $H$.
It is expressed as $\tau_{\mathrm{MT}}={\hbar \pi}/{(2\Delta H)}$, where $(\Delta H)^2=\langle H^2\rangle-\langle H\rangle ^2$ is the variance of the Hamiltonian.
Different from the MT bound, the ML bound characterizes the same problem but relies on the expectation value of the Hamiltonian. Under the assumption that the ground state energy is zero, the ML bound can be expressed as $\tau_{\mathrm{ML}}={\hbar \pi}/{(2\langle H\rangle)}$.
The unified bound combining these two has been proven to be tight \cite{Toffoli2009}, and later works have extended QSLs to non-orthogonal pure and mixed states undergoing both unitary and open quantum dynamics \cite{Giovannetti-Lloyd2003,Giovannetti-Lloyd2004,JonesKok2010pra,Zwierz2012pra,Braunstein1995pra,BRAUNSTEIN1996AOP,Campo2013,Deffner2013,Taddei2013,Mondal2016,Bender,zheng2013observation}.

When discussing QSLs, selecting an appropriate distance measure in state space is crucial.  
While many distance measures exist \cite{Pires-prx}, the Bures angle \cite{Kakutani1948,bures1969extension}, based on fidelity \cite{UHLMANN1976,Jozsa}, is the most commonly used one. It is defined as $\mathcal{L}(\rho,\sigma)=\arccos \sqrt{F(\rho,\sigma)}$, where $F(\rho,\sigma)=\left(\mathrm{Tr} \sqrt{\sqrt{\rho}\sigma\sqrt{\rho}}\right)^2$ represents the fidelity between quantum states $\rho$ and $\sigma$. Using the Bures angle, a unified bound for unitary evolution has been derived as \cite{Giovannetti_Lloy_EPLd2003,Giovannetti-Lloyd2003,Giovannetti-Lloyd2004,Deffner_2017_review}
\begin{equation}
    \tau_\mathcal{L}(\rho,\sigma)=\max\left\{{ \frac{\hbar}{\Delta  H}\mathcal{L(\rho,\sigma)},\frac{2\hbar}{\pi \langle H\rangle}\mathcal{L}^2(\rho,\sigma)}\right\}.
    \label{eq:bures-eq}
\end{equation}
The unified bound of MT bound and ML bound is a special case of this bound when $\mathcal{L}=\pi/2$.
However, QSLs based on the Bures angle perform poorly for mixed states \cite{Campaioli-prl}. On the one hand, they become relatively loose for mixed states. This is evident in the Bloch sphere representation of a qubit system: as both the initial and final states approach the maximally mixed state (center of the Bloch sphere), the Bures angle approaches zero. Consequently, the QSL tends to zero, while the actual evolution time remains finite. On the other hand, from the experimental perspective, obtaining the QSLs based on the Bures angle is highly resource-intensive. This is because the Bures angle is related to the fidelity, the determination of which necessitates quantum state tomography. Given that the number of measurements required for tomography grows exponentially with the number of qubits $n$ \cite{quantum-state-tomography}, this poses a significant practical challenge.

To address these limitations, the Bloch angle is employed. This quantity demonstrates robustness against reductions in state purity and can be experimentally measured through the swap test \cite{swap-test,Cincio_2018,zhan}, a method that requires significantly fewer measurements compared to full tomography. 
In addition, the Bloch angle possesses a clear and intuitive physical interpretation, as it directly corresponds to the rotation angle under unitary evolution. This characteristic makes it more suitable for analyzing unitary dynamics compared to the Bures angle.
Based on the Bloch angle, we propose a novel lower bound for the evolution time required for a quantum system to evolve to a target distance. In contrast to the existing Bloch-angle-based bound which relies on prior knowledge of the actual evolution time \cite{Campaioli-prl}(thus limiting its utility as a lower bound), our QSL does not require this information. Furthermore, it requires fewer computational resources and provides tighter constraints for certain dynamics and initial states. We also demonstrate the measurement of our QSL by experimentally determining the Bloch angle with the swap test in a photonic system.

\section{Theoretical framework}
\label{sec:def}

\begin{figure*}[]
    \centering
\includegraphics[width=\linewidth]{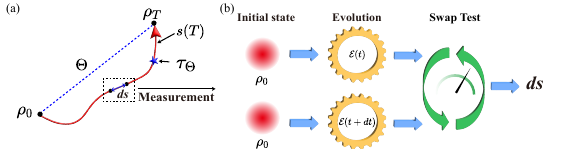}
    \caption{Illustration of a quantum system's evolution path and its measurement. (a) The blue dashed line represents the geodesic line connecting the initial state $\rho_0$ and the final state $\rho_T$, with its length corresponding to the distance $\Theta$ between these two states. The red solid line illustrates the actual trajectory of the quantum system during its evolution. At an intermediate time $\tau_\Theta$ ($0 \leq \tau_\Theta \leq T$), marked by the blue star, the cumulative path length reaches $\Theta$. (b) Procedure for measuring the path length: two identical initial states $\rho_0$ are prepared in separate systems and evolved to times $t$ and $t+dt$ respectively. A swap test is performed to determine the overlap $\text{Tr}(\rho_t\rho_{t+dt})$, allowing the path length $ds$ between $t$ and $t+dt$ to be calculated. Integration of $ds$ yields the total path length at any desired time.}
    \label{fig:schematic}
\end{figure*}

Any density matrix can be expressed as $\rho=(I+\sqrt{{N(N-1)/2}}\vec{r}\cdot\vec{\lambda})/N$ \cite{Nielsen_Chuang_2010},
where $\vec{r}$ is called the Bloch vector, $N$ is the dimension of state space, $I$ is the identity matrix and $\vec{\lambda}$ is a $(N^2-1)$-dimensional vector with components representing the generators of $\mathrm{SU}(N)$. These generators satisfy $\mathrm{Tr}(\lambda_i\lambda_j)=2\delta_{ij}$. The Bloch angle is defined as the angle between two Bloch vectors:
\begin{equation}
\Theta(\rho,\sigma):=\arccos\left(\frac{\vec{r}_1\cdot\vec{r}_2}{|\vec{r}_1| |\vec{r}_2|}\right),
\end{equation}
where $\vec{r}_1,\vec{r}_2$ are the Bloch vectors corresponding to density matrices $\rho,\sigma$, respectively.
Alternatively, an equivalent and computationally simpler formula can be derived as: 
\begin{equation}
    \Theta(\rho,\sigma)=\arccos\frac{N \text{Tr}\left(\rho\sigma\right)-1}{ \sqrt{(N \text{Tr}\left(\rho^2\right)-1)(N \text{Tr}\left(\sigma^2\right)-1)}}.
    \label{eq:bloch angle}
\end{equation}
The reduced sensitivity of the Bloch angle to the decrease of purity is a general property for unital evolutions. A unital evolution is a type of evolution that can preserve the maximally mixed state $I/N$. If two states have parallel initial Bloch vectors (i.e., $\vec{r}'=\eta\vec{r}$ for some scale factor $\eta$), then the two states are related by $\rho'=\eta\rho+(1-\eta) I/N$. Because unital operations preserve $I/N$, applying a unital evolution $\mathcal{C}$ gives $\mathcal{C}(\rho')=\mathcal{C}(\eta \rho)+\mathcal{C}((1-\eta )I/N)=\eta \mathcal{C}(\rho)+(1-\eta) I/N$.
Thus, the proportionality between the Bloch vectors remains invariant: $\vec{r}'(t)=\eta \vec{r}(t)$ for any time $t$ in a unital evolution. This means that the unital evolution of Bloch vectors is independent of purity \textemdash making the Bloch angle effective for application to mixed states.

To derive our QSL, we begin by defining the evolution velocity for unitary evolution. For systems governed by unitary dynamics, the evolution of the density matrix is described by $d\rho_t/dt=-i[H_t,\rho_t]$, where $H_t$ is the Hamiltonian of the system, which is in general time-dependent. We set $\hbar=1$ here and in the remainder of this paper. Considering a state $\rho_{t+dt}$ infinitesimally close to $\rho_t$, we expand it for infinitesimal $dt$ up to second order as $\rho_{t+dt}=\rho_t-i[H_t,\rho_t]dt-i[\partial_t H_t,\rho_t]dt^2/2-\{H_t^2,\rho_t\}dt^2/2+H_t\rho_tH_t dt^2+\mathcal{O}(dt^3).$ Substituting $\rho_t$ and $\rho_{t+dt}$ into Eq. (\ref{eq:bloch angle}), we obtain: 
\begin{equation}
\begin{aligned}
    \Theta(\rho_t,\rho_{t+dt})
    &\approx\sqrt{\frac{2N\mathrm{Tr}[(H_t^2\rho_t^2)-(H_t\rho_t)^2]}{N\mathrm{Tr}(\rho_t^2)-1}} dt.
\end{aligned}
\end{equation}
Since $\Theta(\rho_t,\rho_{t+dt})$ is proportional to $dt$, we define the velocity of evolution under unitary dynamics as:
\begin{equation}        
v(t):=\lim_{dt\to 0}\frac{\Theta(\rho_t,\rho_{t+dt})}{dt}=\sqrt{\frac{2N\mathrm{Tr}[(H_t^2\rho_t^2)-(H_t\rho_t)^2]}{N\mathrm{Tr}(\rho_t^2)-1}}.
\label{eq:expression for velocity}
\end{equation}
This expression quantifies the instantaneous velocity of the system's evolution, forming the basis of our QSL.

According to the triangle inequality, we can establish the following relationship:
\begin{equation}
    \Theta(\rho_0,\rho_T)\le \int_0^{T} \Theta(\rho_t,\rho_{t+dt})=\int_0^{T}v(t)dt\equiv s(T),
    \label{eq:triangle inequality}
\end{equation}
where $s(T)$ represents the path length traversed by the quantum system. Different from $\Theta(\rho_0,\rho_T)$ which relies solely on the initial and final states, $s(T)$ is a path-dependent quantity. When the dynamical path connecting $\rho_0$ and $\rho_T$ follows a geodesic line, $\Theta(\rho_0,\rho_T)=s(T)$. 
Based on this fundamental inequality, two QSLs arise. The first is the existing Bloch-angle-based QSL which is defined as \cite{Campaioli-prl}:
\begin{equation}
\tilde{\tau}_{\Theta}:=\frac{\Theta}{\bar{v}}=\frac{\Theta}{s(T)}T,
\label{eq:def of traditional bound}
\end{equation}
where $T$ is the actual time it takes for a quantum system to evolve a distance $\Theta$ (i.e. $\Theta(\rho_0,\rho_T)=\Theta$) and the average velocity $\bar{v}$ is defined as $s(T)/T$. From Eq. (\ref{eq:triangle inequality}) and Eq. (\ref{eq:def of traditional bound}), we derive that $\tilde{\tau}_\Theta\le T$, with equality achieved when the system's dynamics follow a geodesic line. Notably, determining the existing bound generally requires prior knowledge of the actual time $T$.

\begin{figure}[t]
    \centering
    \includegraphics[width=\linewidth]{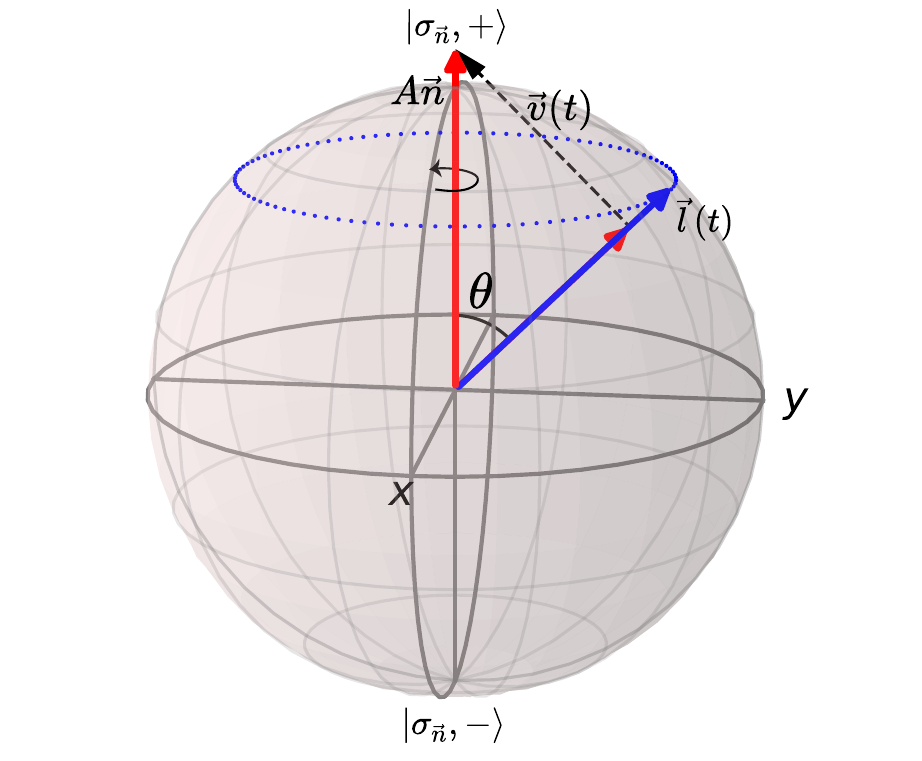}
    \caption{Evolution of a Bloch vector $\vec{l}(t)$ governed by the Hamiltonian $H=A\vec{n}\cdot \vec{\sigma}/2+BI$. The blue dots trace the trajectory of the Bloch vector's evolution. $A\vec{n}$ is the angular velocity for rotation of the Bloch vector and the black dashed vector represents the component of the angular velocity perpendicular to $\vec{l}(t)$ and its length is the  instantaneous evolution velocity $v(t)$. $\theta$ is the angle between the Bloch vector and angular velocity.}
    \label{fig:TI-theo}
\end{figure}

Contrary to the existing bound which is based on the average velocity, our bound is defined as the solution of the integral equation:
\begin{equation}
\Theta=\int_0^{\tau_\Theta}v(t)dt=\int_0^{\tau_\Theta}\Theta(\rho_t,\rho_{t+dt}).
    \label{eq:def of new bound}
\end{equation}
A similar definition based on the Bures angle can be found in \cite{Mirkin-PRA}. Since $\Theta\leq \int_0^T v(t)dt=s(T)$ and $v(t)$ is non-negative, we conclude that $\tau_\Theta$ in our definition is less than or equal to $T$. Moreover, when the path connecting the initial state $\rho_0$ and the 
final state $\rho_T$ follows a geodesic line, $\tau_\Theta=T$. Therefore, the newly defined bound is indeed a lower bound of the actual time $T$, and the condition for it to be equal to the actual time is the same as that of the existing bound. 

In the context of unitary dynamics, where the magnitude of the Bloch vector $\vec{r}$ is preserved throughout the evolution, the geodesic equation is given by
\begin{equation}
\Ddot{\vec{r}} = -\alpha^2 \vec{r},
\end{equation}
with $\alpha$ representing a positive constant equivalent to the instantaneous evolution velocity $v(t)$. The derivation of this equation is provided in Appendix \ref{sec:geodesic-equation}. This scenario is analogous to circular motion in classical mechanics, characterized by acceleration directed toward the center.

\begin{figure*}
    \centering
    \includegraphics[width=\linewidth]{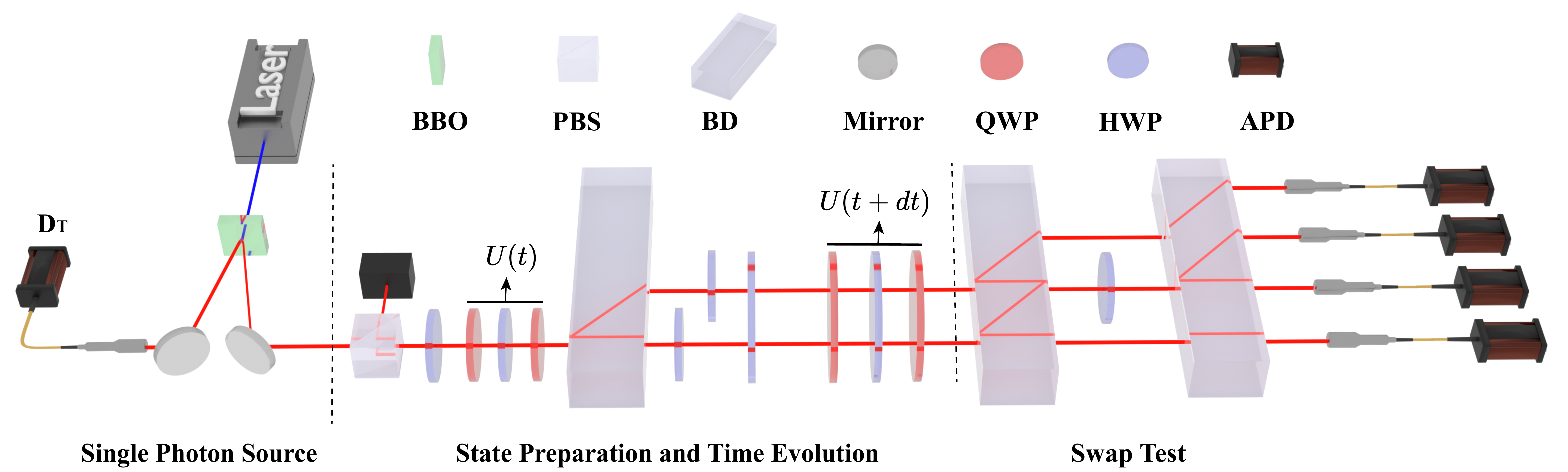}
    \caption{Experimental setup for measurement of quantum speed limit. Photon pairs are generated through the parametric downconversion process by pumping beta-barium-borate (BBO) crystal. One photon is sent to the following setup and is heralded by the detection of the other photon at the heralding detector $D_T$. Two copy of initial states are prepared in photons' polarization modes and path modes respectively by half-wave plates (HWPs) and calcite beam displacers (BDs). The time evolution of this two copies are realized by two combinations composed of two quater-wave plates (QWP) and a HWP. The swap test module include two BDs and a HWP with its optial axis fixed at $22.5^\circ$ relative to horizontal direction.}
    \label{fig:implementation}
\end{figure*}

To elucidate our QSL more clearly, we depict a typical evolution of quantum state in Fig. \ref{fig:schematic} (a). The red line illustrates the path traced out by the quantum system during its evolution. Typically, this path deviates from the geodesic line connecting the initial state $\rho_0$ and the final state $\rho_T$ , resulting in the path length $s(T)$ exceeding the distance $\Theta$ between $\rho_0$ and $\rho_T$. Consequently, there exists an intermediate time $\tau_\Theta \leq T$ at which the system traversed a path length equal to $\Theta$. This $\tau_\Theta$ is our newly defined QSL. When the evolution path coincides with the geodesic line, the inequality $\tau_\Theta \leq T$ becomes an equality.

If we define the average velocity over the interval $[0,\tau_\Theta]$ as $\bar{v}^*=\Theta/\tau_\Theta$, then our QSL represents the time required for quantum states to evolve a distance $\Theta$ at this velocity. In contrast, the existing QSL $\tilde{\tau}_\Theta$ corresponds to the time needed to evolve the same distance at the velocity $\bar{v}$ which is the average velocity over the entire evolution period from $0$ to $T$. Consequently, if the quantum system's evolution accelerates, then $\bar{v}>\bar{v}^*$, our QSL is tighter than the existing QSL. In addition, determining the existing QSL requires solving the dynamical equation or performing measurements over the entire interval $[0,T]$, whereas our bound only involves computations and measurements up to $\tau_\Theta$, thereby reducing computational and experimental resources.

Notably, the new QSL defined here based on the Bloch angle is experimentally accessible. As illustrated in Fig. \ref{fig:schematic} (b), this can be achieved by preparing two identical initial states $\rho_0$ and evolving them to $\rho_t$ and $\rho_{t+dt}$ respectively. The overlap $\text{Tr}(\rho_t\rho_{t+dt})$ can then be measured using the swap test \cite{swap-test-traditional,swap-test,Cincio_2018,zhan} rather than quantum state tomography. Using measured overlaps, the Bloch angle can be readily calculated according to Eq. (\ref{eq:bloch angle}) and the newly defined QSL can be determined according to Eq. (\ref{eq:def of new bound}). Importantly, this measurement framework is highly versatile and applicable to any type of dynamics.

\section{Closed systems with time-independent Hamiltonians}
For closed systems with a time-independent Hamiltonian $H$, the quantum state at time $t$ is determined by $\rho_t=U(t)\rho_0U^\dagger(t)$, where $U(t)=\exp(-iHt)$. In such cases, $\mathrm{Tr}(H^2\rho_t^2)=\mathrm{Tr}(H^2\rho_0^2)$ and $\mathrm{Tr}(H_t\rho_t)^2=\mathrm{Tr}(H\rho_0)^2$, making the velocity $v(t)$ constant according to Eq. (\ref{eq:expression for velocity}) and equal to its initial velocity $v(0)$. Under this condition, calculating the two QSLs is straightforward, as it only requires the initial velocity. Substituting the initial velocity into the expressions of QSLs in Eq. (\ref{eq:def of traditional bound}) and Eq. (\ref{eq:def of new bound}), we find that both $\tau_\Theta$ 
and $\Tilde{\tau}_\Theta$ equal to $\Theta/v(0)$, rendering the two QSLs equivalent. 

As an example, consider a qubit system with a general time-independent Hamiltonian $H= A\vec{n}\cdot \vec{\sigma}/2 +BI$, where $A$ and $B$ are time-independent parameters, $\vec{\sigma}$ denotes a vector with its three components equal to the Pauli operators, $\vec{n}$ is a unit direction vector and $I$ represents the identity matrix. According to the Schrödinger equation, the evolution operator is $U(t)=\exp(-i A\vec{n}\cdot\vec{\sigma}t/2)\exp(-iBt)$ and its action on a spin operator $\vec{\sigma}\cdot\vec{l}$ is to rotate the direction of the spin about the axis $\vec{n}$ through an angle of $At$. Therefore, as shown in Fig. \ref{fig:TI-theo}, when $U(t)$ is applied to a quantum state $\rho=(I+\vec{\sigma}\cdot\vec{l})/2$, the Bloch vector of this state will rotate about $\vec{n}$ with angular velocity $A\vec{n}$. This angular velocity can be decomposed into components parallel and perpendicular to the Bloch vector, but only the perpendicular component induces rotation. Consequently, the evolution velocity is $v(t)=A\sin\theta(t)$, where $\theta(t)$ is the angle between the rotation axis $\vec{n}$ and the Bloch vector $\vec{l}(t)$ at time $t$. For a given Hamiltonian, choosing an initial state with $\theta=\pi/2$ maximizes the evolution velocity to $A$. Under this condition, the Bloch vector evolves along a geodesic, and the two QSLs both become equal to the actual time. To sum up, leveraging the explicit physical interpretation of the Bloch angle, we can easily determine the velocity without calculating Eq. (\ref{eq:expression for velocity}). 

In this work, we utilize a photonic system to simulate unitary dynamics and demonstrate the measurement of QSLs. 
As illustrated in Fig. \ref{fig:implementation}, in our experiment, two identical initial qubit states are encoded in photons' polarization modes (horizontal polarization $|0\rangle$ and vertical polarization $|1\rangle$) and path modes (bottom path mode $|1\rangle$ and top path mode $|0\rangle$). These two states are independently evolved by applying time evolution operators $U(t_1)$ and $U(t_2)$ respectively, which are implemented using a combination of wave plates and calcite beam displacers. The function of a calcite beam displacer is to shift horizontally polarized light while maintaining the path of vertically polarized light. After the evolution, to realize swap test, we perform an orthogonal measurements by projecting the photons onto four basis states:
$|a\rangle \equiv|0\rangle|0\rangle$, $|b\rangle \equiv(|0\rangle|1\rangle+|1\rangle|0\rangle)/\sqrt{2}$, $|c\rangle \equiv (|0\rangle|1\rangle-|1\rangle|0\rangle)/{\sqrt{2}}$, $|d\rangle \equiv|1\rangle|1\rangle$.
As shown in Fig. \ref{fig:implementation}, this measurement is realized using the measurement module which consists of two calcite beam displacers and a half-wave plate with its optical axis fixed at an angle of $22.5^\circ$ relative to the direction of horizontal polarization.
The overlap between $\rho_{t_1}$ and $\rho_{t_2}$ is determined from the probability distribution using the equation: 
\begin{equation}
\text{Tr}(\rho_{t_1}\rho_{t_2})=1-2p_c,
\label{eq:overlap-exp}
\end{equation}
where $p_c$ denotes the probability of obtaining the anti-symmetric state $|c\rangle$ in the measurement. Derivation of Eq. (\ref{eq:overlap-exp}) can be found in Appendix \ref{sec:swap-test}.
Once the overlap is measured, the Bloch angle can be calculated using Eq. (\ref{eq:bloch angle}). By discretizing a time interval $t$ into $N$ small intervals of equal length $\Delta t$, the path length at times $t$ is approximated as $s(t)\approx \sum_{k=0}^{N-1} \Theta(\rho_{k\Delta t},\rho_{(k+1)\Delta t})$.

\begin{figure}[t]
    \centering
    \includegraphics[width=\linewidth]{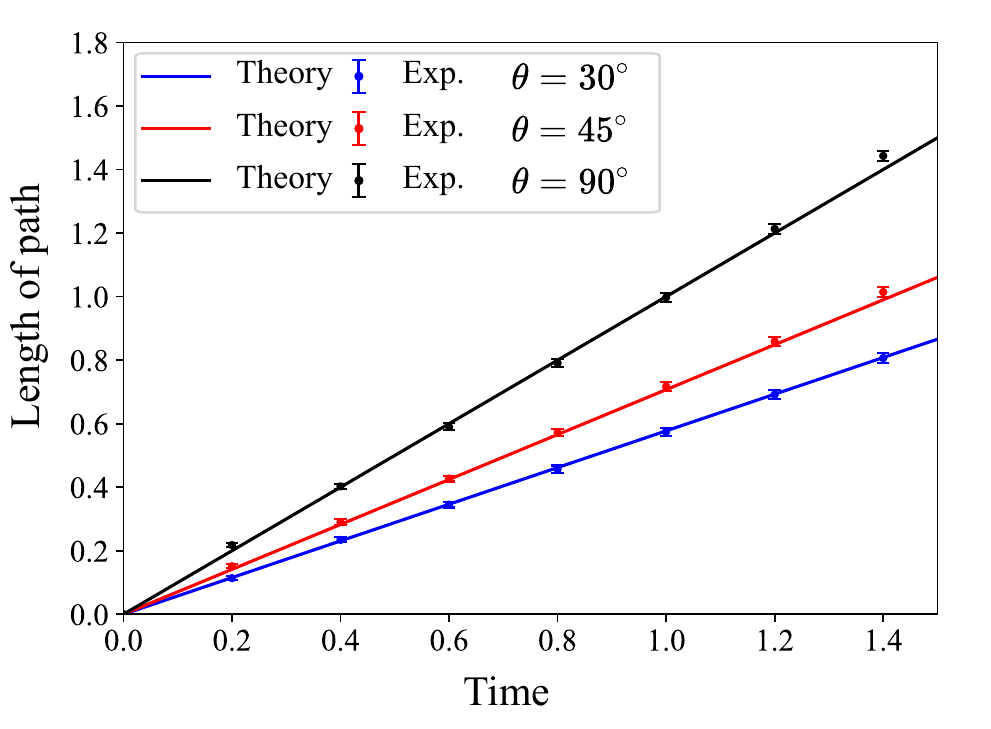}
    \caption{Experimental results of path length as a function of time for a system governed by a time-independent Hamiltonian $H=-1/2\sigma_z$. The three lines represent the evolution from three distinct initial states, each labeled by the relative angle of their Bloch vectors with respect to the rotation axis.}
    \label{fig:TI-exp}
\end{figure}

We demonstrate the evolution of a quantum system governed by a time-independent Hamiltonian $H=-\sigma_z/2$ and the results are shown in Fig. \ref{fig:TI-exp}. In this figure, we see that the relationship between path length and time is a straight line, indicating that the velocity of evolution is constant. When the initial quantum state equals to $(|0\rangle+|1\rangle)/\sqrt{2}$ \textemdash a state whose Bloch vector is perpendicular to the rotation axis within the Bloch sphere \textemdash the evolution velocity reaches its maximum value. The results are in accord with the theoretical predictions within experimental errors. 

\section{Driven systems}
A driven system is governed by unitary dynamics with a time-dependent Hamiltonian. The Landau-Zener model is a typical example of such systems. It describes the transition between the ground state and excited state of a quantum particle as a function of speed and energy gap \cite{landau1932theory,zener1932non}. The Hamiltonian of the Landau-Zener model is given by:
\begin{equation}
    H=Vt\sigma_z+\Delta \sigma_x,
\end{equation}
where $V$ characterizes the speed at which the two energy levels approach each other and $\Delta$ represents the minimal energy gap between the ground and excited states.

The Hamiltonian of the Landau-Zener model can be transformed into $H=A(t)\vec{n}(t)\cdot \vec{\sigma}/2$, where $A(t)=2\sqrt{(Vt)^2+\Delta^2}$ and $\vec{n}=({\Delta}/{(\sqrt{(Vt)^2+\Delta^2})},0,{Vt}/{(\sqrt{(Vt)^2+\Delta^2})})$. Here, $A(t)$ represents the angular velocity, which increases with time. This implies that for most quantum states, the velocity $v(t)$ increases with time, making our QSL tighter than the existing one for most states. Further evidence can be found in Appendix \ref{sec:comparision-for-l-z-model}.

\begin{figure}[t]
    \centering
    \includegraphics[width=\linewidth]{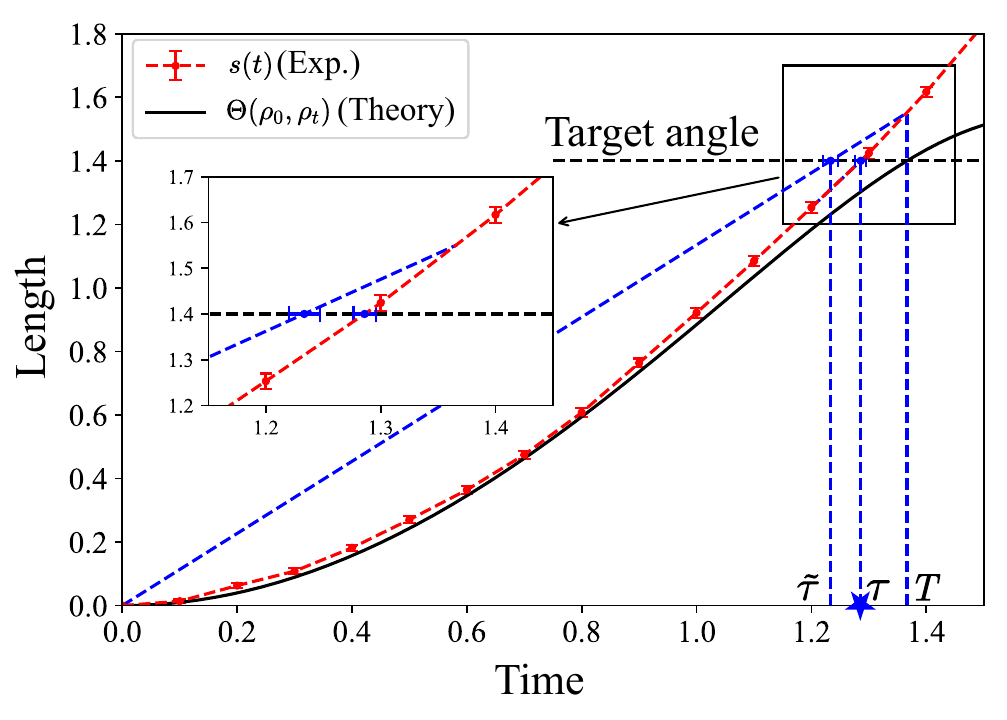}
    \caption{The experimental path length $s(t)$ and theoretical Bloch angle $\Theta(\rho_0,\rho_t)$ as functions of time $t$ for the Landau-Zener model with the initial state equal to $(|0\rangle+|1\rangle)/{\sqrt{2}}$. $\tilde{\tau},\tau,T$ represent the existing QSL, the newly defined QSL and the actual time for the quantum system to evolving to a target angle $1.4$.}
    \label{fig:exp_res_lz}
\end{figure}
In the experiment, we set the parameters in the Landau-Zener model as $V=1$ and $\Delta=1$. 
We choose the state $|\psi\rangle=(|0\rangle+|1\rangle)/{\sqrt{2}}$ as an example to measure both the existing and the new QSLs in our photonic setup.
As shown in Fig. \ref{fig:exp_res_lz}, given a target Bloch angle, we can directly read the new QSL from the graph. The existing QSL can be obtained with the help of some auxiliary lines. The path length as a function of time is convex, reflecting acceleration in Landau-Zener dynamics, which results in our bound being tighter than the existing one. 
It is worth pointing out that the gap between our QSL and the actual time arises from the deviation of the actual path taken by the quantum system from the geodesic line. If the actual path becomes closer to the geodesic line, the constraints imposed by the QSL will become tighter.
Additionally, it can be observed from the figure that determining the existing QSL requires more resources. On one hand, it is necessary to precisely determine the actual evolution time $T$ which is unnecessary for our QSL; on the other hand, the path length for entire evolution process over the interval $[0,T]$ must be measured, which means more measurements need to be performed.

\section{Conclusions}
\label{sec:conclusions}
In this study, we have advanced the understanding of quantum speed limit (QSL) by introducing a new geometric bound based on the Bloch angle. The Bloch angle's robustness against variations of state purity makes it a superior distance measure for mixed state scenarios. Our theoretical framework establishes that the newly defined QSL requires fewer computational resources and offers tighter constraints in accelerated evolution compared to the existing Bloch-angle-based QSL. Experimentally, we validated our theoretical predictions using a photonic system, successfully measuring the Bloch angle through a swap test, thereby demonstrating an efficient method for the determination of Bloch-angle based QSL. This work not only deepens the geometric understanding of QSLs but also paves the way for practical applications in quantum computation, communication, and control. Future research can further explore the application of our Bloch-angle-based QSL in more complex dynamics, such as open quantum systems and imaginary time evolution.

\begin{acknowledgments}
This work was supported by National Natural Science Foundation   of   China (Grants No. 12347104, No. U24A2017, No. 12461160276, No. 12175075), the National Key Research and Development Program of China (Grants No. 2023YFC2205802), Natural Science Foundation of Jiangsu Province (Grants No. BK20243060, BK20233001), in part by State Key Laboratory of Advanced Optical Communication Systems and Networks, China.
\end{acknowledgments}

\appendix

\section{INSTANTANEOUS VELOCITY FOR GENERAL DYNAMICS}
\label{sec:velocity-open}
Consider a quantum system governed by the Lindblad master equation \cite{Nielsen_Chuang_2010}:
\begin{equation}
\begin{aligned}
\frac{d\rho_t}{dt}&=\mathcal{L}(\rho_t)\\
&=-i[{H},{\rho_t}]+\sum_{a>0}\left({L}_a{\rho_t}{L}_a^\dagger-\frac{1}{2}{L}_a^\dagger{L}_a{\rho_t}-\frac{1}{2}{\rho_t}{L}_a^\dagger{L}_a\right),
\end{aligned}
\end{equation}
where $H$ is the Hamiltonian of the system and $L_a$ are quantum jump operators.
For an infinitesimal time $dt$, the state $\rho_{t+dt}$ can be approximated as $\rho_{t+dt}\approx \rho_t+\mathcal{L}(\rho_t)dt$.
Substituting this into Eq. (\ref{eq:bloch angle}), we obtain:
\begin{equation}
\begin{aligned}
&\Theta(\rho_{t+dt},\rho_{t})\\
&=\arccos\frac{N \text{Tr}(\rho_t\rho_{t+dt})-1}{ \sqrt{\left[N \text{Tr}\left(\rho_t^2\right)-1\right]\left[N \text{Tr}(\rho_{t+dt}^2)-1\right]}}\\
&=\arccos\frac{N \text{Tr}\left[\rho_t(\rho_{t}+\mathcal{L}(\rho_{t})dt)\right]-1}{ \sqrt{\left[N \text{Tr}(\rho_t^2)-1\right]\left[N \text{Tr}(\rho_{t}+\mathcal{L}(\rho_{t})dt)^{2}-1\right]}}\\
\end{aligned}
\end{equation}
where $N$ is the dimension of the quantum system.
Let us define the following terms to simplify the expression:
\begin{equation}
\begin{aligned}
f_t&=N\mathrm{Tr}(\rho_{t}^{2})-1,\\
g_{t}&=N\mathrm{Tr}(\rho_{t}\mathcal{L}(\rho_{t})),\\
h_{t}&=N\mathrm{Tr}(\mathcal{L}(\rho_{t}))^{2},
\end{aligned}
\end{equation}
With these definitions, the Bloch angle becomes:
\begin{equation}
\begin{aligned}
\Theta(\rho_{t+dt},\rho_{t})
&=\arccos \frac{f_{t}+g_{t}dt}{\sqrt{ f_{t}(f_{t}+2g_{t}dt+h_{t}dt ^2)}}\\
&=\arccos \frac{1+\frac{g_{t}}{f_{t}}dt}{\sqrt{ 1+2\frac{g_{t}}{f_{t}}dt+\frac{h_{t}}{f_{t}}dt ^2}}\\
&\approx \arccos\left[\left( 1+\frac{g_{t}}{f_{t}}dt \right)\left( 1-\frac{g_{t}}{f_{t}}dt-\frac{1}{2}\frac{h_{t}}{f_{t}}dt ^2 \right)\right]\\
&\approx \arccos\left( 1-\left( \frac{g_{t}}{f_{t}} \right)^{2}dt ^{2}-\frac{1}{2}\frac{h_{t}}{f_{t}}dt ^2 \right)\\
&\approx \sqrt{ 2 \left( \frac{g_{t}}{f_{t}} \right)^{2}+\frac{h_{t}}{f_{t}}}dt.
\end{aligned}
\end{equation}
Since $\Theta(\rho_t,\rho_{t+dt})$ is proportional to $dt$, the instantaneous velocity is therefore:
\begin{equation}
v(t)=\sqrt{ 2 \left( \frac{g_{t}}{f_{t}} \right)^{2}+\frac{h_{t}}{f_{t}}}.
\end{equation}
For unitary dynamics, the Lindbladian simplifies to $\mathcal{L}(\rho_t)=-i[H,\rho_t]$. Under this condition, the term $g_t=-iN[\mathrm{Tr}(\rho_tH\rho_t-\rho_t\rho_tH)]=0$. Consequently, the primary contributions to the instantaneous velocity arise from $h_t$ and $f_t$. For unitary dynamics, $f_t$ is given by
\begin{equation}
\begin{aligned}
h_t&=N\mathrm{Tr}(-i[H,\rho_t])^2\\
&=-N\mathrm{Tr}(H\rho_tH\rho_t-H\rho_t^2H-\rho_t H^2\rho_t-\rho_tH\rho_tH)\\
&=2N\mathrm{Tr}[(H^2\rho_t^2)-(H\rho_t)^2].
\end{aligned}
\end{equation}
Therefore, the instantaneous velocity for unitary dynamics becomes
\begin{equation}
    v(t)=\sqrt{\frac{2N\mathrm{Tr}[(H^2\rho_t^2)-(H\rho_t)^2]}{N\mathrm{Tr}(\rho_t)^2-1}},
\end{equation}
which is the same as Eq. (\ref{eq:expression for velocity}) in the main text.

\section{GEODESIC EQUATION UNDER UNITARY DYNAMICS}
\label{sec:geodesic-equation}
In the main text, we have demonstrated that the geodesic equation under unitary dynamics is represented by 
\begin{equation}
\ddot{\vec{r}} = -\alpha^2 \vec{r},
\label{eq:bloch-geodesic-equation}
\end{equation}
where $\alpha$ denotes a real constant. To ascertain if this equation qualifies as the geodesic equation, it is adequate to verify whether its solution saturates the inequality as stated in Eq. (\ref{eq:triangle inequality}) of the main text.

First, from standard ordinary differential equation theory, the solution of this equation is given by
\begin{equation}
\vec{r}(t) = \cos(\alpha t)\,\vec{r}(0) + \sin(\alpha t)\,\vec{r}'.
\label{eq:solution-to-geodesic-Bloch}
\end{equation}
Under unitary dynamics, the length of the Bloch vector remains constant, which requires that
\begin{equation}
\vec{r}(t)\cdot\dot{\vec{r}}(t)=0.
\end{equation}
Substituting Eq. (\ref{eq:solution-to-geodesic-Bloch}) into the condition yields
\begin{equation}
\begin{aligned}
\vec{r}(t)\cdot\dot{\vec{r}}(t)
&= \frac{1}{2}\alpha\sin(2\alpha t)\Bigl(|\vec{r}(0)|^2 - |\vec{r}'|^2\Bigr)+ \alpha\cos(2\alpha t) \vec{r}(0)\cdot\vec{r}' \\
&= 0.
\end{aligned}
\label{eq:app-ortho}
\end{equation}
Clearly, $\vec{r}'$ in Eq. (\ref{eq:solution-to-geodesic-Bloch}) should satisfy
\begin{equation}
|\vec{r}'| = |\vec{r}(0)|,\quad \vec{r}'\cdot\vec{r}(0)=0.
\end{equation}
Under these conditions, the angle between the Bloch vector at time $T$ and the initial Bloch vector is
\begin{equation}
\begin{aligned}
\Theta(\rho_{0},\rho_{T}) &= \arccos\frac{\vec{r}(T)\cdot\vec{r}(0)}{|\vec{r}(T)||\vec{r}(0)|} \\
&= \arccos\frac{\cos(\alpha T) \vec{r}(0)\cdot\vec{r}(0)}{|\vec{r}(0)||\vec{r}(0)|}\\
&= \alpha T.
\end{aligned}
\end{equation}

To further ascertain the geodesic condition, we need to calculate the path length $s(T)$, defined as the integral of the instantaneous evolution velocity $v(t)$ of the Bloch angle. $v(t)$ can be obtained by computing the magnitude of $\dot{\vec{r}}(t)$. Expanding $\vec{r}(t+dt)$ in a Taylor series up to second order gives
\begin{equation}
\vec{r}(t+dt) \approx \vec{r}(t) + \dot{\vec{r}}(t) dt + \frac{1}{2}\ddot{\vec{r}}(t) dt^2.
\end{equation}
Using Eq. (\ref{eq:app-ortho}), one obtains
\begin{equation}
\vec{r}(t)\cdot\vec{r}(t+dt) \approx |\vec{r}(t)|^2 + \frac{1}{2} \vec{r}(t)\cdot\ddot{\vec{r}}(t) dt^2.
\end{equation}
Then, we have
\begin{equation}
\begin{aligned}
v(t)&=\frac{\Theta(\rho_t,\rho_{t+dt})}{dt}\\
&= \frac{1}{dt}\arccos\frac{\vec{r}(t)\cdot\vec{r}(t+dt)}{|\vec{r}(t)|^2} \\
&= \frac{1}{dt}\arccos\left(1 + \frac{\vec{r}(t)\cdot\ddot{\vec{r}}(t)}{2|\vec{r}(t)|^2} dt^2\right).
\end{aligned}
\end{equation}
Using the approximation $\arccos(1-x)\approx \sqrt{2x}$ for small $x$, we find that
\begin{equation}
\begin{aligned}
v(t)=\frac{1}{dt}\sqrt{-\frac{\vec{r}(t)\cdot\ddot{\vec{r}}(t)}{|\vec{r}(t)|^2} dt^2}.
\end{aligned}
\end{equation}
Using Eq. (\ref{eq:app-ortho}), we can obtain
\begin{equation}
\frac{d}{dt}\bigl(\vec{r}(t)\cdot\dot{\vec{r}}(t)\bigr) = |\dot{\vec{r}}(t)|^2 + \vec{r}(t)\cdot\ddot{\vec{r}}(t)=0.
\end{equation}
Therefore
\begin{equation}
-\vec{r}(t)\cdot\ddot{\vec{r}}(t)=|\dot{\vec{r}}(t)|^2.
\end{equation}
Thus, the instantaneous velocity becomes
\begin{equation}
v(t)=\frac{|\dot{\vec{r}}(t)|}{|\vec{r}(t)|}.
\label{eq:velocity-dot_r}
\end{equation}
Substituting the solution in Eq.~\eqref{eq:solution-to-geodesic-Bloch} leads to
\begin{equation}
\begin{aligned}
v(t)&=\frac{|\dot{\vec{r}}(t)|}{|\vec{r}(t)|}\\
&= \frac{\sqrt{\alpha^2\sin^2(\alpha t)\,\vec{r}(0)\cdot\vec{r}(0)+\alpha^2\cos^2(\alpha t)\,\vec{r}'\cdot\vec{r}'}}{|\vec{r}(0)|}\\
&= \alpha.
\end{aligned}
\end{equation}
Therefore, the total path length is
\begin{equation}
s(T)=\int_{0}^{T}v(t)\,dt = \alpha T = \Theta(\rho_{0},\rho_{T}).
\end{equation}
According to Eq. (\ref{eq:triangle inequality}), the solution of Eq. (\ref{eq:bloch-geodesic-equation}) indeed satisfies the geodesic evolution condition. Thus, for a unitary system, Eq. (\ref{eq:bloch-geodesic-equation}) serves as the geodesic equation.

\section{COMPARISON OF THE EXISTING QSL AND THE NEW QSL IN THE LANDAU-ZENER MODEL}
\label{sec:comparision-for-l-z-model}
\begin{figure}[t]
    \centering
    \includegraphics[width=\linewidth]{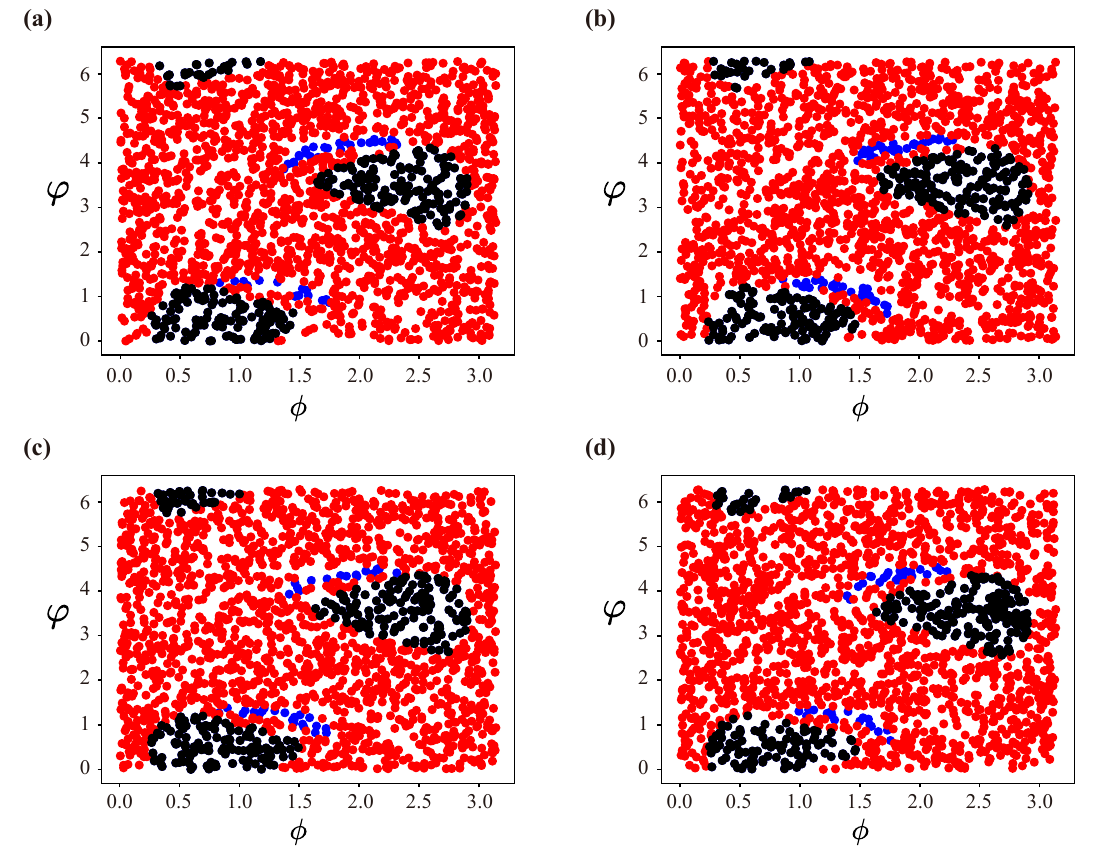}
    \caption{Comparison of the two QSL bounds (the existing QSL $\tilde{\tau}$ and the newly defined QSL $\tau$) for a quantum system governed by the Landau-Zener model. The target Bloch angle is chosen as $\frac{\pi}{2}$ and 2000 initial states are randomly sampled for each purity level: (a) purity=1, (b) purity=0.9, (c) purity=0.8, (d) purity=0.7. The relations between $\phi,\varphi$ and components of Bloch vectors of initial states are $x=r\sin(\phi)\cos(\varphi)$, $y=r\sin(\phi)\sin(\varphi)$ and $z=r\cos(\phi)$. Red, blue points represent $\tau_{\frac{\pi}{2}}>\tilde{\tau}_{\frac{\pi}{2}}$, $\tau_{\frac{\pi}{2}}<\tilde{\tau}_{\frac{\pi}{2}}$ respectively. Initial states which cannot reach the angle $\frac{\pi}{2}$ are denoted by black points. The parameters of the Landau-Zener model are set as $\Delta=1$ and $V=1$.}
    \label{fig:l-z-res}
\end{figure}

To compare the performance of the existing Bloch-angle-based QSL and our new QSL in Landau-Zener model, we randomly sampled 2000 initial states for each purity level. We set the target Bloch angle to $\pi/2$ and compare their tightness via numerical calculation. As illustrated in Fig. \ref{fig:l-z-res}, our new bound outperforms the existing bound for the majority of initial states. Furthermore, the distribution patterns of points remain consistent across varying purity levels, suggesting that the insensitivity of Bloch angle to quantum states' purity in unital evolution. 

\section{SWAP TEST}
\label{sec:swap-test}
Suppose that we have two quantum systems, which we label as $A$ and $B$ and they are in states $|\psi_1\rangle$ and $|\psi_2\rangle$, respectively. In the conventional swap test method \cite{swap-test-traditional}, an ancillary qubit is typically employed. This qubit is initialized in the state $(|0\rangle + |1\rangle)/\sqrt{2}$, 
and is used to control whether the swap operation is applied to the target systems. Specifically, if the ancillary qubit is in the state $|1\rangle$, the swap is applied, while if it is in $|0\rangle$, no swap operation occurs. This procedure generates the state
$(|0\rangle\,|\psi_1\rangle_A|\psi_2\rangle_B+|1\rangle\,|\psi_2\rangle_A|\psi_1\rangle_B)/\sqrt{2}$. 
Subsequently, a Hadamard gate is applied to the ancillary qubit, followed by a measurement. The probability of obtaining the outcome $|1\rangle$ is $(1-|\langle \psi_1|\psi_2\rangle|^2)/2$,
which directly yields the overlap of the two systems.

In fact, the ancillary qubit is not strictly necessary. Here, we derive an alternative scheme without the ancillary qubit. First, we introduce the swap operator $S$, which acts on the composite Hilbert space of systems $A$ and $B$ by exchanging their quantum states. Assume that the bases for the state spaces of $A$ and $B$ are given by $\{ |u_i\rangle \}$ and $\{ |v_j\rangle \}$, respectively. Then, the action of the swap operator on the basis of $A+B$ is given by
\begin{equation}
S|u_i\rangle_A|v_j\rangle_B = |v_j\rangle_A|u_i\rangle_B,\quad \forall\, i,j.
\end{equation}
Thus, ${}_A\langle u_k|\,{}_B\langle v_l|S|u_i\rangle_A|v_j\rangle_B = \langle u_k|v_j\rangle\,\langle v_l|u_i\rangle$. Noting that
\begin{equation}
\begin{aligned}
{}_A\langle u_i|\,{}_B\langle v_j|S^\dagger|u_k\rangle_A|v_l\rangle_B
&= \Bigl({}_A\langle u_k|\,{}_B\langle v_l|S|u_i\rangle_A|v_j\rangle_B\Bigr)^*\\
&= \langle v_j|u_k\rangle\,\langle u_i|v_l\rangle\\
&= {}_A\langle u_i|\,{}_B\langle v_j|S|u_k\rangle_A|v_l\rangle_B,
\end{aligned}
\end{equation}
we can deduce that $S^\dagger = S$, i.e $S$ is a Hermitian operator and can be spectrally decomposed. Moreover, since performing the swap operation twice leaves the system unchanged, it follows that $S^2 = I$.
If $\lambda$ is an eigenvalue of $S$ with corresponding eigenstate $|\psi_\lambda\rangle$, then $S^2|\psi_\lambda\rangle = \lambda^2|\psi_\lambda\rangle = |\psi_\lambda\rangle$, implying that $\lambda^2 = 1$. Therefore, the eigenvalues of $S$ can only be 1 or -1. We denote the eigenstates corresponding to the eigenvalue 1 by $|S_m\rangle$ (the symmetric states) and those corresponding to -1 by $|A_n\rangle$ (the anti-symmetric states), then
\begin{equation}
S\,|S_m\rangle = |S_m\rangle,\quad S\,|A_n\rangle = -|A_n\rangle.
\end{equation}
According to the spectral decomposition, the swap operator can be written as
\begin{equation}
S= \sum_m |S_m\rangle\langle S_m| - \sum_n |A_n\rangle\langle A_n| = P_S - P_A,
\label{eq:S-spectral}
\end{equation}
where $P_S = \sum_m |S_m\rangle\langle S_m|$ and $P_A = \sum_n |A_n\rangle\langle A_n|$
are the projection operators on the symmetric and anti-symmetric subspaces, respectively.

Let the density matrices of the two systems be $\rho_1$ and $\rho_2$, with spectral decompositions
\begin{equation}
\rho_1 = \sum_i p_i\,|w_i\rangle\langle w_i| \quad \text{and} \quad \rho_2 = \sum_j q_j\,|h_j\rangle\langle h_j|.
\end{equation}
The expectation value of $S$ is then given by
\begin{equation}
\begin{aligned}
\mathrm{Tr}(\rho_1\otimes \rho_2 S)
&= \mathrm{Tr}\Biggl(\sum_{i,j} p_i q_j|w_i,h_j\rangle\langle w_i,h_j|S\Biggr)\\
&= \sum_{i,j} p_i q_j\,\langle w_i,h_j|S|w_i,h_j\rangle\\
% & = \sum_{i,j} p_i q_j\,\langle w_i,h_j|h_j,w_i\rangle \\
&= \sum_{i,j} p_i q_j\, |\langle w_i|h_j\rangle|^2.
\end{aligned}
\end{equation}
Since
\begin{equation}
\mathrm{Tr}(\rho_1\rho_2) = \sum_{i,j} p_i q_j\, |\langle w_i|h_j\rangle|^2,
\end{equation}
it immediately follows that
\begin{equation}
\mathrm{Tr}(\rho_1\rho_2) = \mathrm{Tr}(\rho_1\otimes \rho_2S).
\end{equation}
In other words, the overlap between the two systems is equal to the expectation value of the swap operator.

By employing the spectral decomposition of $S$ given in \eqref{eq:S-spectral}, we further obtain
\begin{equation}
\mathrm{Tr}(\rho_1\rho_2)
= \mathrm{Tr}(\rho_1\otimes \rho_2(P_S - P_A))
= p_S - p_A
= 1 - 2p_A,
\end{equation}
where $p_S = \mathrm{Tr}(\rho_1\otimes \rho_2 P_S) \quad \text{and} \quad p_A = \mathrm{Tr}(\rho_1\otimes \rho_2 P_A)$ 
denote the probabilities that the composite system $A+B$ is in the symmetric and anti-symmetric subspaces, respectively. Consequently, by directly measuring the probability of the composite system being in the anti-symmetric state, one can determine the overlap $\mathrm{Tr}(\rho_1\rho_2)$. This is the method adopted in the main text.

\end{document}